\documentclass[12pt]{article}

\usepackage{cite}

\usepackage[nointlimits,reqno]{amsmath}
\usepackage{amssymb}
\numberwithin{equation}{section} 

\usepackage{graphicx}

\usepackage[pdftex,
 pdfproducer=TeXShop,
 pdfcreator=pdflatex]{hyperref}

\usepackage{xspace}
\usepackage{luty}

\newcommand{\ee}{entanglement entropy\xspace}
\newcommand{\MP}{\ensuremath{M_{\text{P}}}\xspace}
\newcommand{\MPb}{\ensuremath{M_{{\text{P}0}}}\xspace}

\begin{document}

\begin{titlepage}

\title{Renormalization of Entanglement Entropy\\
and the Gravitational Effective Action}

\author{Joshua H. Cooperman\ \ and\ \ Markus A. Luty}

\address{Physics Department, University of California, Davis\\
Davis, California 95616}

\begin{abstract}

\end{abstract}
The \ee associated with a spatial boundary in quantum field
theory is UV divergent, with the leading term proportional to the area of
the boundary.
We show that, for a class of quantum states defined by a path integral, 
the Callan-Wilczek formula gives a renormalized geometrical definition of the \ee.
In particular, UV divergences localized on the entangling surface 
do not contribute to the \ee.
The leading contribution to the \ee is then given 
by the Bekenstein-Hawking formula,
and subleading UV-sensitive contributions 
are given in terms of renormalized
couplings of the gravitational effective action.
These results hold even if the UV-divergent contribution
to the entanglement entropy is negative, for example, in theories 
with non-minimal scalar couplings to gravity.
We show that the subleading UV-divergent
terms in the renormalized \ee depend nontrivially on the quantum state.
We compute new subleading terms in the \ee and find
agreement with the Wald entropy formula in all cases.
We speculate that the \ee of an arbitrary spatial boundary
may be a well-defined observable in quantum gravity.
\end{titlepage}

\section{Introduction}
\label{sec:intro}
The discovery of Hawking radiation established that black holes are thermal 
objects \cite{Hawking:1974sw}:
a system containing a black hole obeys the laws of thermodynamics if we associate 
to the black hole an entropy given by the Bekenstein-Hawking formula
(in $D$ spacetime dimensions)
\beq
\eql{SBH}
S_{\rm BH} = \sfrac 14 \MP^{D-2} A_{D-2},
\eeq
where $A_{D-2}$ is the $(D-2)$-dimensional area of the horizon,
and $\MP$ 
is the Planck mass 
\cite{Bekenstein:1973ur,Hawking:1976de,Wald:1999vt}.
It was first suggested by Sorkin \cite{Sorkin:1983aa} that the entropy
of a black hole could be identified with the \ee
\beq\eql{Sent}
S_{\text{ent}} = -\Tr(\rho\ln\rho)
\eeq
associated with the reduced density matrix $\rho$ of the quantum
fields outside the horizon.
Sorkin also pointed out that the leading contribution to this entropy
was UV divergent and proportional to the area.
These ideas were further developed and explicit calculations
of \ee in quantum field theory were carried out in
\Refs{Bombelli:1986rw,Srednicki:1993im}
and for the case of black holes in 
\Ref{Frolov:1993ym}.
It was proposed by Susskind and Uglum \cite{Susskind:1994sm}
that the UV divergences in the area term of the \ee could be absorbed
in the renormalization of the gravitational coupling so that 
the Bekenstein-Hawking entropy of a black hole can be understood
as \ee.
(See also \Ref{Jacobson:1994iw}.)
This led to a large amount of work that appeared to confirm the proposal
in some cases but not in others
\cite{Callan:1994py,Solodukhin:1994yz,Fursaev:1994pq,Fursaev:1994ea,
Demers:1995dq,Kabat:1995eq,Larsen:1995ax,Fursaev:1996uz}. 
We will comment further on the literature after we have stated our
results.

The \ee is defined by a quantum state and an entangling surface, both
of which are  defined on a time slice $\Si$ in spacetime.
If the quantum state is given by a Euclidean path integral, then
the \ee is defined by the geometry of the 
spacetime and the codimension-2 entangling surface.
In this setting there is a beautiful geometric formulation of \ee
due to Callan and Wilczek  \cite{Callan:1994py}.
(Closely related formulas had been proposed earlier for the entropy
of a black hole \cite{Gibbons:1976ue,'tHooft:1984re,Wald:1993nt,
Banados:1993qp,Susskind:1993ws,Carlip:1993sa}.)
The \ee can be written in terms of  the response of the quantum effective
action to a conical singularity at the entangling surface:
\beq
\eql{CWintro}
S_\text{ent} = -\lim_{\de \to 0}
\left( 2\pi \frac{\d}{\d\de} + 1 \right) W_{\text{E},\de},
\eeq
where $\de$ is the deficit angle associated with the conical singularity,
and $W_{\text{E},\de}$ is the Euclidean
quantum effective action in the presence of the conical singularity.
\Eq{CWintro} holds for spacetime geometries with a rotation symmetry
that leaves the entangling surface invariant since only for these
geometries is the conical deficit characterized completely
by a deficit angle $\de$.
In Lorentzian signature these spacetimes have a boost symmetry
that leaves the entangling surface invariant,
that is, a bifurcate Killing horizon
in which the bifurcation surface is the entangling surface.
The Callan-Wilczek formula \Eq{CWintro} is conventionally justified by
continuing $\Tr(\rho^n)$ from integer $n$ (the ``replica trick'').
This is difficult to justify rigorously since there are analytic
functions such as $\sin(n\pi)$ that vanish for all integers.
We give a path integral derivation of the Callan-Wilczek formula 
for rotationally symmetric metrics
that does not rely on the replica trick.

In \Eq{CWintro} $W_{\text{E},\de}$ is the full gravitational effective action, 
including all counterterms required to cancel UV divergences.
\Eq{CWintro} therefore implies that all UV divergences of the \ee are associated
with UV divergences of the gravitational effective action $W_{\text{E},\de}$
on the spacetime with a conical singularity.%
\footnote{In the language of effective field theory, the renormalized
\ee depends on physical UV mass scales, such as the masses of heavy
particles.
The dependence on physical UV mass scales is the same parametrically as the
dependence on the UV cutoff. 
Our discussion is in terms of the UV cutoff because this
is the language used in most of the literature.}
Because this spacetime is singular, $W_{\text{E},\de}$  has
UV divergences that are not present in the gravitational effective action
for smooth spacetimes.
That is, we have the UV-divergent terms
\beq
\eql{generaldivergences}
\begin{split}
W_{\text{E},\de} &= \!\! \int\limits_{\text{spacetime}} 
\!\! \left( c_0 \La^D + c_2 \La^{D-2} R_D + \cdots \right)
\\
&\qquad
+ \!\! \int\limits_{\substack{\text{entangling}\\ 
\text{surface}}} 
\!\! \left( c'_0 \La^{D-2} + c'_2 \La^{D-4} R_{D-2} + \cdots \right)
+ \cdots,
\end{split}
\eeq
where $\La$ is the UV cutoff, 
$R_D$ and $R_{D-2}$ are the Ricci scalars of the $D$-dimensional spacetime
metric and the $(D-2)$-dimensional induced metric on the entangling surface,
and $\La^0$ is understood to mean $\ln \La$.
We can think of the entangling surface as a codimension-2 brane
in spacetime, and the additional UV-divergent terms localized on 
the brane are a 
consequence of the fact that such a brane is a UV modification of the theory.
One of the main results of this paper is that the UV divergences of the 
\ee are 
nonetheless independent of the brane-localized UV-divergent terms.
The reason is that the latter arise only at $O(\de^2)$ 
while the \ee depends only on the $O(\de)$ terms.
These results are established 
using a careful regularization of the singular conical spacetime.

The leading UV-divergent term in the \ee arises from the Einstein-Hilbert term 
$c_2 \La^{D-2} R_D$ in the gravitational effective action.
This generates the UV-divergent area term in the \ee:
\beq
\eql{arearesult}
S_\text{ent} = -4\pi c_2 \La^{D-2} A_{D-2} + \cdots.
\eeq
The coefficient of $R(g)$ in the Euclidean gravitational effective action
is $-\MP^{D-2} / 16\pi$, so \Eq{arearesult}
is precisely $+\frac 14$ times the UV-divergent contribution to
$\MP^{D-2} A_{D - 2}$.
More generally, for any $D$-dimensional local term in the gravitational 
effective action, there is a corresponding contribution to the \ee,
and we give an algorithm for computing it.
These results hold for any spacetime dimension, to all
orders in perturbation theory, and for all subleading as well as leading
UV divergences.
They hold for a general quantum field theory
coupled to a background metric but 
not for quantum fluctuations of the metric itself.
This restriction arises because we do not have
a satisfactory generally covariant
regulator for the conical singularity in the presence of quantum 
fluctuations of gravity.

Another restriction is that the results are established
only for the special class of spacetimes discussed above.
This is equivalent to considering a special class of quantum states.
In order for the spacetime without the deficit angle 
to be non-singular, the entangling
surface must have vanishing extrinsic curvature in the time
slice $\Si$.
These restrictions mean that we cannot treat some cases
of interest, such as the vacuum state with a nontrivial entangling surface.
For black holes the only quantum state to which our methods apply
is the Hartle-Hawking state.
These limitations are closely related to the problem of defining
\ee for a general spacetime metric.
Generalizing our methods to overcome these 
restrictions is an important open problem.


The area term is independent of the quantum state of the system, but
we show by explicit calculation that
the subleading UV-divergent terms in the \ee depend nontrivially
on the quantum state.
This can be seen from the fact that these subleading terms
depend on geometrical invariants that are
not intrinsic to the time slice $\Si$ on which the quantum state
is defined.
The spacetime geometry away from $\Si$ determines the quantum
state, so this represents dependence on the quantum state.
It is a familiar feature of quantum field theory that subleading
UV divergences can depend on infrared physics.
For example, in the presence of a particle with mass $m$,
the cosmological constant in $D=4$ spacetime dimensions
will have UV-divergent 
contributions of the form 
$\sim \La^4 + m^2 \La^{2} + m^4 \ln \La$.
It seems that this dependence of subleading UV-divergent terms in the \ee
on the quantum state has not been
appreciated in the literature.

When we add local $D$-dimensional counterterms to cancel the UV divergences
in the gravitational effective action, the results above imply that
\Eq{CWintro} gives a finite result for the \ee.
The leading area term in the \ee is then given by the
Bekenstein-Hawking formula
\beq
\eql{SBHrenormintro}
S_\text{ent} = \sfrac 14 \MP^{D-2} A_{D-2} + \cdots,
\eeq
where $\MP$ is the renormalized Planck scale.
If the quantum field theory is an effective theory obtained by 
matching to some more fundamental theory above the cutoff $\La$, 
then the counterterms are determined by requiring that the predictions
of the effective theory agree with those of the fundamental theory.
Physical quantities are independent of $\La$ in the effective theory
simply  because $\La$ is an arbitrary matching scale.
The corresponding counterterms for the \ee are therefore similarly
interpreted as contributions to the \ee from
correlations of the modes above the cutoff $\La$.

Our interpretation that \Eq{CWintro} gives a renormalized \ee 
removes the objections
raised in the literature to the identification of black hole
entropy with the \ee of the horizon.
In most of the literature, the divergent part of the \ee is identified with
the \ee.
For a physical regulator such as a lattice, the regulated
theory is a unitary quantum system, and the UV-divergent \ee
has a state-counting interpretation; 
however, for applications involving gravity (for example, black holes),
one must use a generally covariant regulator such as Pauli-Villars
or heat kernel regularization,
and the UV-divergent term in the \ee
does not have a sensible state-counting interpretation.
For example, in scalar field theory the UV-divergent contribution
to the \ee depends on the 
curvature coupling $\sfrac 12 \xi R(g) \Phi^2$ and is negative for some values of $\xi$
\cite{Fursaev:1994pq,Demers:1995dq,Larsen:1995ax,Solodukhin:1995ak,
Barvinsky:1995dp}.
In theories with vector fields, the \ee is negative
due to unphysical ``surface contributions'' 
\cite{Kabat:1995eq,Larsen:1995ax,Barvinsky:1995dp,Fursaev:1996uz}.
(For gravitational fluctuations the absence of a satisfactory regulator 
for the conical singularity  does not  permit an unambiguous result for the \ee
\cite{Fursaev:1996uz,Iellici:1996gv}.)
The unphysical features of the UV-divergent \ee
have led to attempts to distinguish between `statistical' and `conical'
definitions of entropy (see for example, 
\Refs{Frolov:1998vs,Donnelly:2012st}).

We instead interpret \Eq{CWintro} as giving a definition of a renormalized \ee.
This formula has no manifest state-counting interpretation, but as we
argued above, neither does the UV-divergent part in covariant regulators.
We will see below that the resulting renormalized entropy agrees
with Wald entropy for black hole spacetimes, providing evidence
{\it a fortiori} that \Eq{CWintro}
is a physically meaningful definition of
entropy.
The renormalized \ee is manifestly generally covariant
and always positive since the leading 
area term is proportional to the renormalized Planck scale.
The physical interpretation is that the 
renormalized \ee includes counterterms
that account for the correlations of modes above the cutoff $\La$.%
\footnote{%
A closely related Wilsonian definition of the \ee\ has been
discussed in \Ref{Jacobson:2012ek}.}

If the entangling surface is the horizon of a black hole, then the area term
in the \ee is the Bekenstein-Hawking entropy, which is 
the leading contribution to the thermodynamic entropy of the black hole.
It is natural to ask whether the subleading terms in the
renormalized \ee for black holes are also physically meaningful. 
We therefore compare the renormalized \ee
with the Wald entropy formula for a black hole in a gravitational theory with
higher-dimension interaction terms in the action \cite{Wald:1993nt}.
The Wald entropy is the thermodynamic entropy for classical dynamics
governed by the gravitational effective action.
The comparison between \ee and Wald entropy therefore makes sense
when the gravitational effective action is obtained by integrating
out heavy modes, and the only massless mode is gravity itself.
In this case the long-wavelength dynamics of the black hole
are governed by the gravitational effective action in a derivative
expansion.
Previous results found agreement between the \ee
and the Wald entropy for terms in the effective action that are algebraic
functions of the Riemann tensor \cite{Jacobson:1994qe,Fursaev:1995ef}.
We compute contributions to the \ee arising from gravitational
interaction terms of the form $(\nabla_\mu R_{\nu\rho\si\tau})^2$,
and we again find agreement.

Finally, we offer some speculations based on the results above.
With some important limitations, we have
established that the gravitational effective action defines
a renormalized \ee.
The limitations are that the result does not apply to fluctuations
of gravity itself and only holds for special classes of entangling surfaces and of 
quantum states.
We find it plausible that our results can be generalized to remove
these limitations.
If this proves to be the case, then
it would suggest that \ee is a well-defined
observable in a complete theory of quantum gravity for any entangling surface,
with the leading contribution given by the Bekenstein-Hawking
formula.%
\footnote{We thank R. Myers for encouraging us to think about
the interpretation of our result for general spacetimes.}
It is believed that, in a complete theory of quantum gravity,
there is a minimum length that can be physically probed.
Entanglement entropy is UV divergent in quantum field theory
due to the presence of correlated modes with arbitrarily short wavelengths.
In a theory with a fundamental length, it is therefore natural for the
\ee to be finite.
Further evidence for this point of view comes from the holographic
\ee formula of Ryu and Takayanagi \cite{Ryu:2006bv},
which applies to entangling surfaces that are more general than
black hole horizons.
On the other hand, the concept of spacetime (and hence of a spacetime boundary)
is presumably an emergent concept in a theory of quantum gravity.
Even in perturbative string theory it is not clear how to define
an entangling surface without introducing physical
states on the surface (for example, $D$-branes).
The generalized conjecture formulated above 
can be studied in spacetime geometries much simpler than that of a black hole, 
for example, flat spacetime with a planar entangling surface.
Further work on this question is clearly motivated.
This conjecture has also been 
discussed in 
\Refs{Fursaev:2006ng,Fursaev:2007sg,Fursaev:2010ix,Bianchi:2012ev}.

The remainder of this paper is organized as follows.
In \S\ref{sec:cone} we give a general discussion of the \ee
in quantum field theory in a gravitational background
and identify the geometries and quantum states for which 
the \ee is given by the Callan-Wilczek formula.
In \S\ref{sec:proof} we discuss the regularization of the conical
singularity and prove our main result.
In \S\ref{sec:conclude} we discuss the implications and limitations
of our results and suggest directions 
for future work.

\section{Entanglement Entropy and Conical Spaces
\label{sec:cone}}
We begin with a discussion of \ee in a general
quantum field theory in a background spacetime geometry.
We identify spacetime geometries and quantum states for which we can 
justify the Callan-Wilczek formula,
thereby giving a geometric renormalized definition of the \ee.

\subsection{Geometrical Formulation
\label{sec:geometrystate}}
\begin{figure}[t]
\begin{center}
\includegraphics[scale=1]{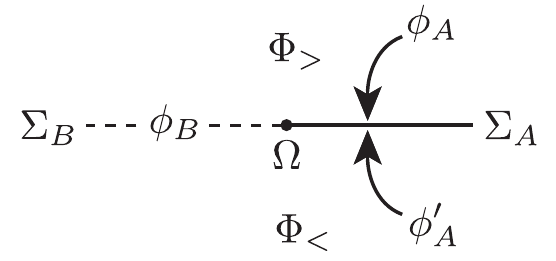} 
\begin{minipage}{5.3in}
\caption{Definition of fields for the 
Euclidean path integral defining the density matrix 
\Eq{rhoreduced} and the quantum states
\Eqs{statepathintket} and \eq{statepathintbra}.}
\label{fig:fields}
\end{minipage}
\end{center}
\end{figure}

The \ee is defined within a quantum field theory for 
a time slice $\Si$, a quantum
state on
$\Si$, and an entangling surface 
$\Om$ that divides $\Si$ into two parts $\Si_A$ and $\Si_B$.
We are interested in the reduced density matrix $\rho_A$ that
describes correlation functions of fields on $\Si_A$.
We can give a geometrical definition of $\rho_A$ using a
path integral for a special class of quantum states on $\Si$.
We denote the spacetime quantum fields by $\Phi$ and their
restriction to the time slice $\Si_{A,B}$ by $\phi_{A,B}$.
The correlation functions of the fields $\phi_A$ is then given
by a path integral (continued to Euclidean time)
\beq
\avg{\phi_{A1} \cdots \phi_{An}}
= \myint d[\Phi] \, e^{-S_{\text E}[\Phi]} \, \phi_{A1} \cdots \phi_{A2}
= \Tr(\rho_A \gap \phi_{A1} \cdots \phi_{A2}),
\eeq
where
\beq\eql{rhoreduced}
\bra{\phi'_A} \rho_A \ket{\phi_A}
= \myint \,\, d[\phi_B]
\!\!\!\!\! \int\limits_{\substack{\Phi(0-) = (\phi_B, \phi_A)
\\ 
\Phi(0+) = (\phi_B, \phi'_A)}}
\!\!\!\!\!
d[\Phi] \, e^{-S_\text{E}[\Phi]}.
\eeq
That is, the density matrix is defined by performing the path
integral over  fields in all of spacetime except $\Si_A$,
with suitable boundary conditions on $\Phi$ above and below
$\Si_A$. (See Fig.~\ref{fig:fields}.)

We now consider the conditions under which this path integral 
computes the reduced density matrix $\rho_{A}$ in a pure quantum state on $\Si$.
We define a quantum state $\ket\Psi$ on $\Si$ by
\begin{align}
\braket{\phi}{\Psi} 
&= \lim_{\ep \to 0} \lim_{T \to \infty} \myint d[\phi_i]\,
\bra{\phi} U(0, -T(1 + i\ep)) \ket{\phi_i}
\\
\eql{statepathintket}
&= \!\!\!\! \int\limits_{\Phi_<(\tau = 0) = \phi} 
\!\!\!\! d[\Phi_<]\, e^{-S_{\text{E}}[\Phi_<]},
\end{align}
where the time slice $\Si$ is at $\tau = 0$.
The path integral is over fields $\Phi_<$ defined
for $\tau < 0$. (See Fig.~\ref{fig:fields}.)
Similarly, we can define a ket state $\bra{\tilde\Psi}$ by
\begin{align}
\braket{\tilde\Psi}{\phi}
&= \lim_{\ep \to 0} \lim_{T \to \infty} \myint d[\phi_f]\,
\bra{\phi_f} U(T(1 + i\ep), 0) \ket{\phi}
\\
\eql{statepathintbra}
&= \!\!\!\! \int\limits_{\Phi_>(\tau = 0) = \phi} 
\!\!\!\! d[\Phi_>]\, 
e^{-S_{\text{E}}[\Phi_>]}.
\end{align}
If $\ket\Psi = \ket{\tilde\Psi}$, then the path integral
\Eq{rhoreduced} computes the reduced density matrix corresponding
to the pure state $\ket\Psi$.
This follows if 
\beq\eql{Ucond}
U(T,0) = U^\dagger(0,-T) = U(-T, 0),
\eeq
which requires the metric to have a reflection symmetry
about $t = 0$.
One can treat more general time slices and quantum states in the path integral
using the Schwinger-Keldysh formalism \cite{Schwinger:1960qe,Keldysh:1964ud},
but we will not discuss that here.

\subsection{Entanglement Entropy from Conical Geometry
\label{sec:conicalee}}
We now turn to the \ee
\beq
S_{\rm ent} = -\Tr(\rho_A \ln\rho_A)
\eeq
associated with the reduced density matrix $\rho_A$ given by \Eq{rhoreduced}.
We show how to derive the Callan-Wilczek formula for this 
entropy for a large class of spacetimes.

We begin with the simplest case, that of flat spacetime with
a planar boundary.
We write the metric in Euclidean space as
\beq
ds_\text{E}^2 = d\tau^2 + dz^2 + \de_{ij} dy^i dy^j,
\eeq
where $\de_{ij}$ is a flat metric for the remaining $D - 2$ directions. 
The time slice $\Si$ is the $\tau = 0$ surface, and the entangling
surface is at $z = 0$.
We can write this as
\beq
ds_\text{E}^2 = dr^2 + r^2 d\th^2 + \de_{ij} dy^i dy^j.
\eeq
where
$\tau = r \sin\th$ and $z = r\cos\th$.
The path integral in \Eq{rhoreduced} can be 
thought of as summing over complete sets of field configurations
on a sequence of half-planes labelled by $\th$. 
We are thus using $\th$ as a Euclidean time variable.
The Hamiltonian $K$ generating evolution in $\th$ is then 
the generator of rotations in $\th$.
Because the system is invariant under translations in $\th$, 
$K$ is independent of $\th$, and we have
\beq\eql{BWresult}
\rho_A = e^{-2\pi K}.
\eeq
The preceding argument follows the discussion of \Ref{Kabat:1994vj}.

These results have a well-known physical interpretation when
continued back to Minkowski spacetime.
Taking $\th \to i\eta$ gives
the flat spacetime metric in the form
\beq
ds^2 = -r^2 d\eta^2 + dr^2 + \de_{ij} dy^i dy^j.
\eeq
The Hamiltonian $K$ now generates translations in $\eta$,
which are boosts about the entangling surface $r = 0$.
The reduced density matrix is therefore thermal with Hamiltonian 
given by the boost generator in Minkowski spacetime
\cite{Bisognano:1975ih,Bisognano:1976za}.
For constant acceleration observers traveling on trajectories
of constant $r$ and $y^i$, the boost parameter is proper time,
so these observers see a thermal
excitation of the quantum field theory, the Unruh effect \cite{Unruh:1976db}.

We can now write the \ee as \cite{Callan:1994py}
\beq
S_{\text{ent}} = \lim_{\ep \to 0} \left( \frac{\d}{\d\ep} + 1 \right)
\ln \Tr(\rho^{1 - \ep})
= \ln \Tr(\rho) - \frac{\Tr(\rho \ln \rho)}{\Tr(\rho)}.
\eeq
The \rhs\ is equal to the \ee for $\Tr(\rho) = 1$
and is independent of rescaling of $\rho$,
which is equivalent to a rescaling of the Euclidean
path integral measure.
We can therefore compute $\Tr(\rho^{1 - \ep})$
by a Euclidean path integral in which
field configurations at $\th = 0$ and $\th = 2\pi(1 - \ep)$ are
identified.
This is equivalent to the Euclidean path integral for the theory
in which the metric has a conical singularity at the origin with a
deficit angle $\de = 2\pi \ep$.
We then have
\beq
\Tr(\rho^{1 - \ep})
= \Tr e^{-(2\pi - \de) K} = 
\myint d[\Phi] \gap\gap
e^{-S_{\text{E},\de}[\Phi]} = e^{-W_{\text{E},\de}},
\eeq
where $W_{\text{E},\de}$ is the Euclidean
effective action on the conical space.
This gives
\beq\eql{CW}
S_{\text{ent}} = - \lim_{\de \to 0} \left( 2\pi \frac{\d}{\d\de} + 1
\right) W_{\text{E},\de},
\eeq
which is the formula of Callan and Wilczek.
The conventional derivation of this result uses the analytic
continuation of $\Tr(\rho^n)$ to non-integer $n$ (the ``replica trick'').
The present discussion gives a derivation that avoids the need for this 
continuation.
The result is, however, formal because the
Hamiltonian $K$ is singular at $r = 0$.
Correspondingly, the path integral for $W_{\text{E},\de}$ is over a space with a conical
singularity at $r = 0$ that requires regularization in addition
to the usual UV regularization of the quantum field theory.
This will be discussed in detail below.

\begin{figure}[t]
\begin{center}
\includegraphics[scale=1]{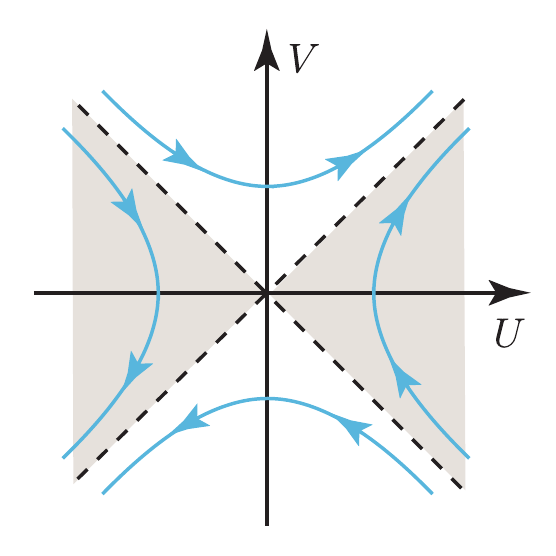} 
\begin{minipage}{5.3in}
\caption{Coordinates for boost invariant spacetime.
The arrows show the orbits of the boost symmetry, and the
shaded region corresponds to $\kappa = U^2 - V^2 > 0$.}
\label{fig:Kruskal}
\end{minipage}
\end{center}
\end{figure}

The preceding discussion can be
generalized to spacetime metrics with a boost symmetry about the
entangling surface.%
\footnote{We thank R. Myers for pointing out the importance
of the rotational/boost symmetry in this context.}
For such a spacetime we can write the metric in the Kruskal-like form
\beq
\eql{Kruskallike}
ds^2 = \om^2(\ka, y) \left( -dV^2 + dU^2 \right) + \ga_{ij}(\ka, y) dy^i dy^j,
\eeq
where 
\beq
\ka = U^2 - V^2.
\eeq
The entangling surface is at $U = V = 0$, and $(V, U)$ transforms as
a Lorentz vector under boosts.
This metric has a bifurcate Killing horizon for the boost symmetry
at $V = \pm U$. (See Fig.~\ref{fig:Kruskal}.)
This class of metrics includes many nontrivial spacetimes of interest,
such as black hole spacetimes and de Sitter space.

We can continue this metric to Euclidean space by writing
$V \to iT$ and defining
\beq
U = R \cos\th,
\qquad
V = R \sin\th,
\eeq
where
\beq
R = \sqrt{\ka} = \sqrt{U^2 + T^2} \ge 0.
\eeq
The resulting Euclidean metric is then
\beq\eql{resultingmetric}
ds_\text{E}^2 = \om^2(R^2, y) (dR^2 + R^2 d\th^2)
+ \ga_{ij}(R^2, y) dy^i dy^j,
\eeq
where the entangling surface is at $R = 0$.
We find it more convenient to write the metric as
\beq\eql{genmetric}
ds^2_\text{E} = dr^2 + \rho^2(r, y) d\th^2
+ \ga_{ij}(r, y) dy^i dy^j.
\eeq
For each $\th = \text{constant}$ slice we are using 
Gaussian normal coordinates $(r, y)$ for the 
entangling surface at $r = 0$.
The function $\rho(r, y)$ gives the circumference of the $\th$ orbit
that passes through the point $(r, y)$ on a $\th = \text{constant}$ slice.
Gaussian normal coordinates may break down far from the
$r = 0$ surface, but we will see that the UV-divergent contributions to the
\ee are sensitive only to the structure of the
spacetime geometry near $r = 0$.

In order for the metric \Eq{genmetric} to be nonsingular at
the entangling surface,
we must have $\rho(r,y) \sim r$ as $r \to 0$.
To write the
conditions on $\rho(r,y)$, 
it is convenient to define
\beq
\rho(r,y) = r \si(r, y).
\eeq
The conditions are then
\begin{gather} 
\eql{alphacondrzero}
\left. \si \right| = 1,
\qquad
\left. \d_r^m \si \right| = 0,
\ \ m = 1, 3, 5, \ldots,
\\
\eql{gammaacondrzero}
\left. \d_r^n \ga_{ij} \right| = 0,
\ \ n = 1, 3, 5, \ldots,
\end{gather}
where $|$ denotes evaluation at $r = 0$.
Note that these conditions hold for arbitrary $y$ so that for example,
 $\d_i \si | = 0$.
The extrinsic curvature tensor of the entangling surface in the
time slice $\Si$ is $K_{ij} = \d_r \ga_{ij} |$, so we see that
this is required to vanish.
The need for the higher $r$ derivatives to
vanish can be understood from requiring
that $\Box^p R(g)$ is nonsingular at $r = 0$ for all $p$.

The derivation of the Callan-Wilczek formula \Eq{CW}
proceeds exactly as above
for the more general metric \Eq{genmetric} since the rotation symmetry 
guarantees that the Hamiltonian $K$ generating rotations in $\th$
is independent of $\th$.
If there is no rotational symmetry, then we can define $K$ by
\Eq{BWresult}, but then $K$ is a non-local operator, and 
$\rho^{1-\ep}$ cannot be computed by simply restricting the range of
the angular ``time'' evolution.

\subsection{Flat Spacetime
\label{sec:groundstate}}
The path integral in flat spacetime defines the vacuum state, which
is a very natural quantum state to study.
The \ee in the vacuum
has a nontrivial dependence on the geometry of the
entangling surface that gives an interesting observable for
general studies of quantum field theory.
For example, for conformal field theories the logarithmically divergent terms 
in the \ee for spherical entangling surfaces in $D = 4$ are
related to conformal anomalies 
\cite{Solodukhin:2008dh,Casini:2010kt,Casini:2011kv}.
However, boosts in flat spacetime can only leave invariant a flat plane,
so the framework described
above can only describe a trivial entangling surface in the vacuum
state.

\subsection{Global Schwarzschild Spacetime
\label{sec:BHspacetime}}
Another interesting special case is the
quantum state defined by a path integral in the maximally
extended Schwarzschild solution.
We illustrate this for $D = 4$,
where the metric is given in Kruskal-Szekeres coordinates by
\newcommand{\RS}{R_\text{S}}
\newcommand{\rs}{r_\text{S}}
\beq
\eql{Kruskal}
ds^2 = \frac{4 \RS}{\rs} e^{-\rs/\RS} \left(
-dV^2 + dU^2 \right) + \rs^2 d\Om^2,
\eeq
where $d\Om^2$ is the metric of $S^2$,
$\RS = 2GM$ is the Schwarzschild radius, and
$\rs$ is the standard Schwarzschild radial coordinate,
given in these coordinates by
\beq
U^2 - V^2 = \left( \frac{\rs}{\RS} - 1 \right) e^{\rs / \RS}.
\eeq
Continuing to Euclidean space and writing the metric in the
form of \Eq{resultingmetric}, we obtain 
\beq
\eql{KruskalE2}
ds_\text{E}^2 = \frac{4 \RS}{\rs} e^{-\rs/\RS} \left(
R^2 d\th^2 + dR^2 \right) + \rs^2 d\Om^2.
\eeq
We can change coordinates to put this in the form
\beq
\eql{ESchwartzschild}
ds^2_\text{E} = \al^2(r) dr^2 + r^2 d\th^2 
+ \RS^2 \al(r) d\Om^2,
\eeq
where
\beq
\al(r) = \left( \frac{4\RS^2}{4\RS^2 - r^2} \right)^2.
\eeq
These coordinates are different from those in \Eq{genmetric},
but they allow a simple explicit form of the metric.
In these coordinates spatial infinity is at $r = 2\RS$,
and the Euclidean ``time'' $\th$ is compact with a finite period
$4 \pi \RS$ at spatial infinity.
The Euclidean path integral in this space therefore defines a thermal state
with the Hawking temperature $T_\text{H} = 1/4\pi \RS$ at infinity, 
the Hartle-Hawking state.
This is in accordance with the general result that any quantum state that is
non-singular at the horizon must have thermal radiation at infinity.

The metric \Eq{ESchwartzschild} is equivalent to the standard
Euclidean Schwarzschild metric obtained by continuing
$t \to i\tau$ in standard Schwarzschild coordinates; 
however, the discussion here clarifies a number of points 
in the standard treatment.
In our discussion the 
Euclidean metric includes the time slice $V = 0$ in physical
spacetime, and it is clear that a path integral in this Euclidean space computes
correlation functions of fields on this slice.
Also, the periodicity in $\th$ is not imposed by hand but arises
from the fact that the spacetime metric is smooth at $U = V = 0$.

From the point of view of the path integral, there is no need for the
spacetime geometry away from the time slice $\Si$ to satisfy the
equations of motion.
What makes the quantum state
defined by the Euclidean Schwarzschild metric special
is that it is invariant under the time translation symmetry 
that corresponds to the boost symmetry in the $(V,U)$ plane.
This is the symmetry that makes the black hole static,
so this is the natural thermal state.
Other spacetime metrics that give the same induced metric
on $\Si$ define different quantum states that can be studied
using path integral methods.

\section{Entanglement Entropy and the Gravitational Action
\label{sec:proof}}
In the previous section we showed that, for spacetimes of the form
\Eq{genmetric}, the \ee can be computed
from the gravitational effective action on a conical space
using the Callan-Wilczek formula \Eq{CW}.
The conical space is, however, singular at $r = 0$ because the Hamiltonian 
$K$ becomes singular there, so we must regulate the conical singularity to
define \Eq{CW}.
This is a UV regularization in addition to the usual
UV regularization of short-distance modes of the quantum fields.
In this section we give a careful discussion of the regularization
of the conical singularity and use it to show that renormalizing
the UV divergences of the gravitational effective action for non-singular metrics
are sufficient to renormalize the \ee.
This has been demonstrated in \Refs{Fursaev:1994ea,Fursaev:1995ef} 
for terms in the gravitational effective
action that are algebraic functions of 
curvature tensors.
The present analysis extends these results to arbitrary terms in the 
effective action and gives a simple universal result for the corresponding
contribution to the \ee.

\subsection{Regulating the Cone}
We begin by describing the regulator for the conical space that we use.
Regulated conical spaces were discussed in \Ref{Fursaev:1995ef},
but we use a different regulator to prove results for UV-divergent
terms in any spacetime dimension and at any order in the derivative expansion.
For the general metric \Eq{genmetric} we make the replacement
\beq\eql{genmetriccone}
ds^2_{\text{E}} \to d\tilde{s}^2_{\text{E}} =
dr^2 + \rho^2(r, y) \left[1 - \ep \be(r) \right]^2 d\th^2
+ \ga_{ij}(r, y) dy^i dy^j,
\eeq
where 
\beq
\eql{thebeta}
\be(r) = \Th_+(r) = \lim_{\ell \to 0+} \Th(r - \ell).
\eeq
Derivatives of $\Th_+$ are distributions localized at
the coordinate endpoint $r = 0$, so the limit $\ell \to 0+$
is needed to define it precisely.
In the metric \Eq{genmetriccone} the circumference of a small circle
of radius $r$ is $2\pi (1 - \ep) r$, so this describes a space
with deficit angle $\de = 2\pi \ep$.

If we keep $\ell \ne 0$, then the metric \Eq{genmetriccone} is not continuous
at $r = \ell$, so it may be objected that this is not
a fully regulated metric.
Only for smooth background metrics are we guaranteed that the UV divergences
in the gravitational effective action are given by local $D$-dimensional
terms.
We define a smooth regulated metric by replacing
$\Th(r - \ell)$ 
in \Eq{thebeta} by a smooth
step function that varies on the scale $\ell' \ll \ell$.
We then take the limit $\ell' \to 0$ followed by $\ell \to 0$ to remove
the regulator.

The fully smoothed metric gives the same result in this limit as the
distribution \Eq{thebeta} because the \ee depends on the terms in the gravitational effective
action that are linear in $\ep$.
These terms consist of one power of (derivatives of) $\be(r)$
multiplied by a smooth function, which is well-defined.
This gives the same result as the limit $\ell' \to 0$,
$\ell \to 0$ in the fully smoothed metric.
If we go beyond linear order in $\ep$, then the $\ell' \to 0$
limit is singular because derivatives give terms of order $1/\ell'$
that diverge as $\ell' \to 0$.%
\footnote{To obtain the $\La$ dependence of the \ee we must
take the limit $\ell' \to 0$, $\ell \to 0$ with $\La$ held fixed.
This is particularly clear in an effective field theory 
where the cutoff scale is identified with the physical mass of a
heavy particle.}
These represent additional UV divergences in the effective action
that can be cancelled by counterterms localized on the singular surface
of the form \Eq{generaldivergences}.

We write the UV-divergent terms in the gravitational effective
action as
\beq
W_{\rm E,\de} = \myint d^D x \gap \sqrt{\tilde{g}}\, \scr{F}(\tilde{g}), 
\eeq
where $\scr{F}(\tilde{g})$ is a sum of local invariants constructed
from the regulated metric and its derivatives.
We then have
\begin{align}
S_{\rm ent} &= -\lim_{\ep \to 0}
\left( \frac{\d}{\d\ep} + 1 \right)
\myint d^D x\gap \sqrt{\tilde{g}}\, \scr{F}(\tilde{g})
\nonumber\\
\eql{Sentsofar}
&= -\lim_{\ep \to 0}
\myint d^D x\gap \sqrt{g} \, 
\frac{\d}{\d\ep} \scr{F}(\tilde{g}). 
\end{align}
We see that the \ee depends only on the $O(\ep)$ 
terms in the geometrical invariant $\scr{F}$.

\subsection{The Conical Limit
\label{sec:conicallimit}}
We now carefully consider the $\ell \to 0$ limit of \Eq{Sentsofar}.
We show that an arbitrary curvature invariant
$\scr{F}$ yields
a contribution to the \ee given by an
integral over the entangling surface of a well-defined
geometrical invariant constructed from $\scr{F}$.

In order to work with well-defined tensor quantities, we 
write the unperturbed metric as
\beq
g_{\mu\nu} = n_\mu n_\nu + \xi_\mu \xi_\nu
+ \ga_{\mu\nu},
\eeq
where 
\begin{gather}
\eql{vrhodefn}
n_r = 1, 
\qquad 
n_\th = 0,
\qquad
n_i = 0,
\\
\xi_r = 0, 
\qquad 
\xi_\th = \rho,
\qquad
\xi_i = 0,
\end{gather}
and $\ga_{\mu\nu}$ is nonzero only for the $i, j$ components.
Here, $\xi^\mu$ is the Killing vector associated with the rotational
symmetry about the entangling surface and satisfies the Killing equation
\beq
\nabla_\mu \xi_\nu + \nabla_\nu \xi_\mu = 0.
\eeq

The perturbed metric can then be written as 
$\tilde{g}_{\mu\nu} = g_{\mu\nu} + h_{\mu\nu}$
with 
\beq
h_{\mu\nu} = -2\ep \be \xi_\mu \xi_\nu + O(\ep^2).
\eeq
This shows that $\be$ is a scalar with respect to the unperturbed metric.
We can therefore write \Eq{Sentsofar} in terms of covariant
derivatives of $\be$:
\beq
\eql{SIalmost}
S_\text{ent} = -\myint d^{D} x \gap\gap \sqrt{g}
\gap\gap \sum_{n = 1}^\infty
\scr{F}^{\mu_1 \cdots \mu_n} \nabla_{\mu_1} \cdots \nabla_{\mu_n} \be.
\eeq
The $n = 0$ term with no derivatives acting on $\be$
is absent in \Eq{SIalmost}
because for $\be = \text{constant}$
the perturbation is a rescaling of
the $\th$ coordinate, which does not affect the value of the
invariant $\scr{F}$.
(In the $r$ integral with $\be = \text{constant}$, 
we are effectively integrating 
over the space with $r = 0$ removed.)
We can then integrate \Eq{SIalmost} by parts to write it as an integral
over the first derivative of $\be$:
\beq
S_{\rm ent} = -\myint d^{D} x \gap\gap \sqrt{g}
\gap\gap \tilde{\scr{F}}^\mu \nabla_\mu \be,
\eeq
where
\beq
\tilde{\scr{F}}^\mu = \scr{F}^\mu - \nabla_\nu \scr{F}^{\mu\nu} + \cdots.
\eeq
Using $\nabla_\mu \be = n_\mu \be'$
(with $\be' = \d_r \be$) we have in our coordinates
\begin{align}
S_\text{ent} 
&= -2\pi \myint d^{D-2} y \gap\gap \sqrt{\ga}
\int_0^{r_\infty} dr \gap\gap \rho \gap\gap \be'(r) \gap\gap I[\scr{F}]
\eql{Irint}
\\
\eql{SI}
&= -2\pi \myint d^{D-2} y \gap\gap \sqrt{\ga} \,
\lim_{r \to 0} \rho \gap\gap I[\scr{F}],
\end{align}
where
\beq
I[\scr{F}] = n_\mu \tilde{\scr{F}}^\mu.
\eeq
We have used $\be(r) = \Th_+(r)$ only in the last step of \Eq{SI}.
To see that the $r \to 0$ limit in \Eq{SI} is well-defined,
note that the $r$ integral in \Eq{Irint} must converge 
at $r = 0$ if we replace $\be$ by a smooth function with
$\be' = \text{constant}$.
We can expand $I[\scr{F}]$ in a power series in $r$,
so this implies that
\beq
I(r) = \frac{I_1}{r} + I_0 + O(r).
\eeq
Because $\rho(r,y) = r + O(r^3)$, we have our final result
\beq
S_{\text{ent}} = -2\pi \myint d^{D-2} y \gap\gap \sqrt{\ga} \gap\gap
I_1[\scr{F}].
\eeq
This gives a general algorithm for computing the entanglement entropy,
which can be summarized as follows.
We define the invariant $I[\scr{F}]$ by writing the $O(\ep)$ term in
$\scr{F}$ as $\be'(r)I[\scr{F}]$ using integration by parts.
We then expand $I[\scr{F}]$ in powers of $r$, and the entanglement
entropy density is given by $-2\pi$ times the $1/r$ term in $I[\scr{F}]$.

The fact that the \ee can be computed from the gravitational effective
action without additional UV divergences localized on the conical
singularity is one of the main results of this paper.
Let us reiterate the logic of the argument.
We first replace the singular conical metric with a 
smooth metric  for which all UV divergences
in the gravitational effective action are associated with
local $D$-dimensional terms.
We then consider the limit in which we recover the singular metric,
and we show that the terms contributing to the entanglement
entropy are well-defined and finite.
This demonstrates 
that no additional counterterms are required
to define the \ee.

\subsection{Calculations}
We now perform some calculations using the results above.
The leading UV-divergent term gives rise to the Bekenstein-Hawking
entropy and is independent of the quantum state.
We show that the subleading UV-divergent
terms in the \ee depend nontrivially on the
quantum state.

The perturbed metric \Eq{genmetriccone} is obtained by making
the replacement $\rho \to (1 - \ep \be) \rho$ in \Eq{genmetric},
so we can compute all curvature invariants from this metric.
The nonzero components of the Riemann tensor are
\begin{align}
R_{r\th r\th} &= -\rho\rho''
\\
R_{r\th\th i}&= \rho\d_i \rho'-\sfrac 12 \rho\ga'_{ij}\ga^{jk}\d_k \rho
\\
R_{\th i \th j} &= -\sfrac 12 \rho\rho' \ga'_{ij}-\rho\nabla^{(\ga)}_i\d_j\rho,
\\
R_{r i r j} &=-\sfrac 12 \ga''_{ij} + \sfrac 14 \ga^{k\ell} \ga'_{ik} \ga'_{j\ell},
\\
R_{rijk}&=-\sfrac 12\nabla^{(\ga)}_{j}\ga'_{ki}+\sfrac 12\nabla^{(\ga)}_{k}\ga'_{ij}\\
R_{ijk\ell} &= R_{ijk\ell}(\ga) 
- \sfrac 14 \left[ \ga'_{ik} \ga'_{j\ell} - \ga'_{i\ell} \ga'_{jk} \right]
\end{align}
and those that can be obtained
from the ones above using the symmetries of the Riemann tensor.
Here, a prime denotes differentiation with respect to $r$, and
$\nabla^{(\ga)}_i$ is the covariant derivative with
respect to the metric $\ga_{ij}$.

We use these result to compute the contribution to the \ee arising
from various terms in the gravitational effective action.
We first consider the UV-divergent Einstein-Hilbert term in the Euclidean
gravitational effective action in $D$ dimensions:
\beq
W_{\text{E}} = \myint d^D x \gap \sqrt{g}\, c_2 \gap \La^{D-2} \gap R(g) + \cdots
\eeq
The Ricci scalar in the regulated
metric \Eq{genmetriccone} is 
\beq
R(\tilde{g}) = R(g) + \ep \left[
\frac{4 \rho'\be'}{\rho} +\be'\ga^{ij}\ga'_{ij}+2\be'' \right]
+O(\ep^2).
\eeq
Following the procedure derived in the previous subsection,
we obtain
\beq
I_1[R] = 2.
\eeq
Here, we used the conditions \Eqs{alphacondrzero} and \eq{gammaacondrzero}
to expand the solution about $r = 0$.
We then obtain
\beq
\De S_{\text{ent}} = -2\pi \myint d^{D-2}y \gap \sqrt{\ga}\,
2 \gap c_2 \La^{D-2} = -4 \pi c_2 \La^{D-2} A_{D-2},
\eeq
where $A_{D-2} = \int \! d^{D-2} y \gap \sqrt{\ga}$ is the area of the 
entangling surface.

To obtain a finite gravitational effective action, we add to the action
the counterterm
\beq
\De W_{\text{E}} = -\frac{\MPb^{D-2}}{16\pi} \myint d^D x \gap \sqrt{g} \gap R(g),
\eeq
where $\MPb$ is the bare Planck mass.
As discussed in the introduction, this is interpreted as parameterizing
the contribution of the modes above the cutoff.
The renormalized Planck mass is then given by
\beq
\MP^{D-2} = \MPb^{D-2} - 16\pi c_2 \La^{D-2}.
\eeq
In this case the contribution to the \ee from
the Einstein-Hilbert term is finite and given by the 
renormalized Bekenstein-Hawking formula
\beq
S_\text{ent} = +\sfrac 14 \MP^{D-2} A_{D-2} + \cdots
\eeq

We now consider the subleading UV-divergent terms in
the gravitational effective action:
\beq
\eql{Wgravdiverge}
\De W_{\text{E}} = \myint d^D x \gap \sqrt{g} \gap \left[
 c_{4,1} \La^{D-4} R^2(g)
+ c_{4,2} \La^{D-4} R_{\mu\nu}^2
+ c_{4,3} \La^{D-4} R_{\mu\nu\rho\si}^2 
+ \cdots \right].
\eeq
We use the above results
to compute the invariant $I_1[\scr{F}]$ for the 
curvature-squared invariants, obtaining
\begin{align}
\eql{RiemannsquaredI1}
I_1[R_{\mu\nu\rho\si}^2] &= -8 \rho^{(3)},
\\
\eql{RiccitensorsquaredI1}
I_1[R_{\mu\nu}^2] &= -2 \left[ 2\rho^{(3)} + \ga^{ij}\ga''_{ij} \right],
\\
\eql{RicciscalarsquaredI1}
I_1[R^2] &= -8 \left[ \rho^{(3)} + \ga^{ij}\ga''_{ij}- \sfrac 12 R(\ga) \right].
\end{align}
It should be remembered that the right-hand sides of these equations are evaluated
at $r = 0$.
These results agree with Eqs.~(3.27)--(3.29) of \Ref{Fursaev:1995ef},
which were computed with a different regulator for the conical space.%
\footnote{Another check of these results is that, for the Euler 
term $E_4 = R_{\mu\nu\rho\si}^2 - 4 R_{\mu\nu}^2 + R^2$ in
$D = 4$, we obtain
\beq
I_1[E_4] = 4 R(\ga).
\eeq
$E_4$ is a topological term, and the topology of the
manifold is $R^2 \times X$, so the Euler density must vanish if
$X = R^2$.
$I_1[E_4]$ must therefore
be a $D=2$ topological term, which is indeed the case.}
This calculation can be generalized to 
an arbitrary function 
containing no covariant derivatives acting on the Riemann tensor.
The result can be written in the covariant form
\beq
\eql{I1FR}
I_1[\scr{F}(R_{\mu\nu\rho\si})]
= \left. \frac{\d\scr{F}}{\d R_{\mu\nu\rho\si}}
\left( P_{\mu\rho} P_{\nu\si} - P_{\mu\si} P_{\nu\rho} \right)
\right|_{r = 0},
\eeq
where $P_{\mu\nu}$ is the metric in the space perpendicular to
the entangling surface, that is, 
\beq
P_{rr} = 1,
\qquad
P_{\th\th} = \rho^2,
\eeq
with all other components vanishing.
In using this relation it is important to take into 
account the symmetries of the Riemann tensor so that for example,
\beq
\frac{\d R}{\d R_{\mu\nu\rho\si}} = \sfrac 12 \left(
g^{\mu\rho} g^{\nu\si} - g^{\mu\si} g^{\nu\rho} \right).
\eeq

It is easily seen that the results 
\Eqs{RiemannsquaredI1}--\eq{RicciscalarsquaredI1}
for the subleading UV-divergent 
terms in the \ee cannot be expressed in terms of
the intrinsic geometry of the time slice $\Si$,
which is independent of $\rho(r,y)$.
The \ee is defined by the geometry of $\Si$,
the entangling surface in $\Si$, and the quantum state.
The quantum state is determined by a path integral and therefore
depends on the full spacetime geometry.
The dependence of the \ee
on geometrical invariants that are not intrinsic to $\Si$
therefore represents dependence on the quantum state.
We conclude that the subleading UV-divergent terms in the \ee
depend nontrivially on the quantum state.

The fact that the subleading UV divergences depend on low-energy
quantities should not be surprising.
Just from dimensional analysis, subleading UV divergences can
depend on IR mass scales.
For example, the cosmological constant in $D = 4$ has the
UV-divergent contributions
$\sim \La^4 + \La^2 m^2 + m^4 \ln \La$, 
where $m$ is the mass of a particle.
As this example shows, it is only the leading UV divergence that is 
expected to be independent of IR scales.

Based on these considerations, we expect that the area term in the \ee does not
depend on the quantum state.
The results above show that this is indeed the case
for those quantum states that can be obtained from a path
integral in the class of spacetimes described in \S\ref{sec:conicalee}.
We expect this universality of the area term to hold much more generally.
The leading UV divergence of the \ee arises from the growth of the
density of eigenvalues of $\rho_{A}$ at short wavelengths, and we expect 
the leading behavior of this to be independent of the quantum state
as long as the state does not involve excitations at arbitrarily
short wavelengths.

\subsection{Wald Entropy}
It is interesting to compare the renormalized \ee computed here with
the general entropy formula of Wald \cite{Wald:1993nt}.
The Wald entropy formula holds for a gravitational theory
with arbitrary higher-dimension interaction terms
and for metrics with a bifurcate Killing horizon,
precisely the setup for which the \ee is given by the
Callan-Wilczek formula.
The Wald entropy formula additionally requires that the metric is a stationary
point of the gravitational effective action, while the \ee
does not require this.
Wald entropy is a thermodynamic entropy in the sense
that the (classical) laws of black hole thermodynamics
hold for this entropy.
Agreement between Wald entropy and \ee is therefore
an indication that \ee explains the
thermodynamic entropy of black holes.

It is known that, if the gravitational effective action is an
algebraic function of the Riemann tensor, \ee and Wald
entropy agree for general metrics \cite{Jacobson:1994qe,Fursaev:1995ef}.
Our results allow us to extend this comparison to terms that
involve derivatives of the Riemann tensor.

It makes sense to compare \ee and Wald entropy when
the gravitational effective action is obtained by integrating out heavy
particles, and the only massless degrees of freedom are those of the metric.
In this case the terms in the action with additional derivatives parameterize
small corrections that are treated perturbatively in a derivative
expansion.
The lowest-order terms in the derivative expansion that involve
derivatives of the Riemann tensor are $O(\d^6)$ terms of the form
$(\nabla R)^2$.
As an example we consider the term
\beq\eql{dRiemann2}
\De W_{\text{E}} = \myint d^D x \gap\gap \sqrt{g} \gap\gap
(\nabla_\mu R_{\nu\rho\si\tau})^2
\eeq
for the spherically symmetric metric 
\beq
\eql{specialDD}
ds_{\text{E}}^2 = dr^2 + \rho^2(r) d\th^2
+ \chi^2(r) d\Om_{D-2}^2,
\eeq
where $d\Om_{D-2}^2$ is the metric for the $(D-2)$-dimensional
sphere.
We find that
\beq
\De S_{\text{Wald}} = \De S_{\text{ent}}
= 2\pi \myint d^{D-2} y \gap \sqrt{\ga}\gap\gap\gap
\frac{16}{3} \left[ \left( \rho^{(3)} \right)^2 - \rho^{(5)} \right].
\eeq
The Wald entropy and \ee from \Eq{dRiemann2} agree
for any $D$.
A general argument that \ee and Wald entropy
agree for spherically symmetric metrics has been given in 
\Ref{Nelson:1994na}.

\subsection{Gravitational Fluctuations}
An important limitation of the results above is that
they do not hold for fluctuations of gravity itself.
The problem is that the regulated metric \Eq{genmetriccone} does not 
satisfy the vacuum Einstein equations; therefore, the action for 
the metric fluctuations is not well-defined.
We can think of the unperturbed metric as the solution of the Einstein 
equations with a nontrivial stress-energy tensor.
Consistently extending this to include fluctuations of gravity requires
that the stress-energy tensor be covariantly conserved in the
presence of gravitational fluctuations.
If it is not, then we cannot decouple the unphysical polarizations of
the metric fluctuations.
%
The only known way of satisfying this is for the stress-energy tensor
to be associated with a dynamical theory coupled to gravity.
The conical singularity can be induced by a codimension-2 brane
at $r = 0$; however, this object has massless fluctuations, so it is not a
purely UV modification of the theory.
It may be that these can be decoupled in the limit $\ep \to 0$,
but this analysis is beyond the scope of this paper.


\section{Conclusions
\label{sec:conclude}}
In this paper we have shown that \ee has a renormalized
geometrical definition for a
class of quantum states defined by a path integral in a spacetime with a boost
symmetry about the entangling surface.
For this class of quantum states, the UV divergences in the \ee are in one-to-one
correspondence with the UV divergences in the gravitational effective
action, and renormalizing this effective action gives a renormalized
\ee.
The leading term for large entangling surfaces is 
given by the Bekenstein-Hawking formula 
$\sfrac 14 \MP^{D-2} A_{D-2}$.
These results hold for a general quantum field theory coupled to gravity
in any spacetime dimension and
to all orders in perturbation theory.
We also show that the subleading UV-divergent terms
in the \ee depend nontrivially on the quantum state,
while the leading term is independent of the state.

We argue that the renormalized \ee defined by the renormalized
effective action for gravity is the physical \ee.
The counterterms parameterize
the contribution to the \ee
from modes above the cutoff.
This interpretation removes many of the objections to the
identification of \ee with black hole entropy.

We compared our results for \ee with the Wald entropy formula for
black holes 
in theories with higher derivative terms in the gravitational effective
action.
We found that the $O(\d^6)$ contribution to the \ee
from a gravitational interaction $(\nabla_\mu R_{\nu\rho\si\tau})^2$
agrees with the Wald entropy formula.

The results of this paper have several important limitations.
They do not apply to quantum fluctuations of gravity itself since
in that case we do not have a regulator of the conical singularity 
that preserves general covariance
and does not introduce additional massless degrees of freedom,
thereby modifying the theory in the IR as well as the UV.
Another limitation is the one already mentioned: our results 
have been demonstrated only for a special class of quantum states.
Both of these limitations are related to the problem of finding
a geometrical formulation of \ee for general spacetimes.
We believe it is a very interesting open problem to 
understand the renormalization of the \ee in general gravitational
backgrounds including fluctuations of gravity.

We believe that it is highly plausible that the restriction to
metrics with a boost invariance about the entangling surface
can be removed by a generalization of the present analysis.
The \ee for quantum states defined by a path integral in a general
spacetime is a completely geometrical object, so it is natural
to expect that it can be renormalized by adding counterterms to the
gravitational effective action.
In particular, the UV-divergent terms in the \ee are local
to the entangling surface, and any such surface and the surrounding
geometry are locally flat.
We can therefore introduce curvature perturbatively, and 
it seems reasonable that an analysis of these perturbations
will not destroy the structure we have found in the symmetric
case.
We leave investigation of this question to future work.

\section*{Acknowledgements}
We would like to thank R. Myers for numerous discussions and advice
on many points
and for valuable comments on an early draft.
We also thank S. Carlip, N. Kaloper,
and N. Tanahashi for useful conversations and comments.
We thank D.~Fursaev, T.~Jacobson, J. Maldacena, S.~Solodukhin,  
L.~Susskind, and A.~Wall
for critical comments on the first version of this manuscript.
MAL acknowledges the support of the Jensen Prize
from the Institute for Theoretical Physics at the
University of Heidelberg, where part of this work was carried out.
This research was supported in part by the Department of Energy under 
grant DE-FG02-91ER40674.

\bibliographystyle{utphys}
\bibliography{mycites}

\end{document}